\documentclass[aps,prd,preprint]{revtex4}
\bibliography{apsrev}
\begin{document}

\title{Isotropic representation of noncommutative $2D$ harmonic 
oscillator}

\author{Anais Smailagic}
\email{anais@ictp.trieste.it}
\affiliation{INFN, Sezione di Trieste}
\author{Euro Spallucci}
\email{spallucci@trieste.infn.it}
\affiliation{Dipartimento di Fisica Teorica,
				  Universit\`a di Trieste
		          and INFN, Sezione di Trieste}
\date{\today}

\begin{abstract}
We show that  $2D$ noncommutative harmonic oscillator has an isotropic
representation in terms of commutative coordinates.
The noncommutativity in the new mode, induces energy level 
splitting, and is equivalent to an external magnetic field effect. The
equivalence of the spectra of the isotropic and anisotropic representation
is traced back to the existence of  $SU(2)$ invariance of the noncommutative
model.
\end{abstract}

\pacs{03.65G, 03.65F, 02.40G, 11.15-q}
\maketitle
\section{Introduction}
Recent results obtained in the framework of non-perturbative string theory
\cite{witten},\cite{sw}, have boosted the interest for a deeper 
understanding of the role played by noncommutative geometry in different 
sectors of theoretical physics \cite{schw}. 
Inclusion of noncommutativity in quantum field theory 
can be achieved in two different ways: via Moyal $\ast$-product on the space of
ordinary functions, or defining the field
theory on a coordinate operator space which is intrinsically noncommutative
\cite{nekra}, \cite{presn}. The equivalence between the two approaches has been
nicely described in \cite{agw}. While formally well defined,
the operator approach is hard to implement  in explicit calculations. The
analysis of the noncommutative effects is usually performed expanding
Moyal $\ast$-product  perturbatively, and take into account additional vertices.
In order to get
deeper understanding of the way in which noncommutativity affects quantum
field theory one tries to understand these effects firstly in exactly solvable
models  of Noncommutative Quantum Mechanics \cite{qm}. \\
The difficulty of performing explicit calculation encountered in  operator
space formulation of quantum field theory corresponds, in quantum mechanics,
to the problem  of  formulating a Schroedinger equation directly in terms
of noncommutative coordinates. The path to follow is to introduce the
noncommutativity of coordinates and momenta
 through the Moyal $\ast$-product \cite{greci}. It turns out that the effect of
 introducing the  $\ast$-product can be described by suitable shifts 
 of the argument of the wavefunction \cite{gamb}, or of the Hamiltonian
 \cite{jell}. In order to properly treat the noncommutative variables
 one needs two  commuting Heisenberg algebras \cite{nair}, \cite{bellucci}. \\
In this paper we shall follow an approach where the set of 
noncommutative coordinates $x_i\ , p_i$ is expressed as a linear 
combination of canonical variables of quantum mechanics $\alpha_i\ , \beta_i$.
As we shall see, the noncommutativity will manifest itself through
redefinitions of various parameters and will produce additional 
terms in the Hamiltonian of the equivalent commutative description.\\
 As an explicit example we shall study
 the case of a $2D$ noncommutative harmonic oscillator.
The main result of our work is the description of the noncommutative system
in terms of {\it new} set of transformations among noncommutative and
canonical variables, that we shall name the 
 ``{\it isotropic  representation}''. In this mode, the noncommutative $2D$ 
 harmonic oscillator receives a simple and clear physical  interpretation. 
 This representation also exhibits  the rotational symmetry and leads, in 
 a simple way, to the  
 form of the generator of rotations for the noncommutative representation.
 Finally, we shall explain the equivalence of the spectra in  two different 
 representations in terms of an $SU(2)$ symmetry.\\

\section{Canonical coordinates}
In order to illustrate the general procedure we start with
the set of coordinates and momenta satisfying extended commutators as 
\cite{uno}

 \begin{eqnarray}
&&\left[\, x_k\ , x_j\,\right]=i\,\Theta_{kj}\label{comm1}\\
&&\left[\, p_k\ , p_j\,\right]=i\,B_{kj}\label{comm2}\\
&&\left[\, x^k\ , p_j\,\right]=i\,\delta^k{}_j\label{comm3}
\end{eqnarray}

with $\Theta_{kj}$ and $B_{kj}$ antisymmetric matrices characterizing
generalized noncommutativity of the phase space geometry.\\
We are  going to define  linear transformations 
from the noncommutative set of coordinates $\left(\, x_i\ , p_i\,\right)$ to a 
{\it commutative} set of canonically conjugate coordinates  
$\left(\, \alpha_i\ , \beta_i\,\right)$ which obey

\begin{eqnarray}
&&\left[\, \alpha_k\ , \alpha_j\,\right]=0\label{comm4}\\
&&\left[\,\beta_k\ ,\beta_i  \,\right]=0\label{comm5}\\
&&\left[\,\alpha_k \ ,\beta_i \,\right]=i\,\delta^k{}_j\label{comm6}
\end{eqnarray}

The relation of noncommutative coordinates to conjugate ones is given by

\begin{eqnarray}
&& x_i= a_{\, ij}\, \alpha_j + b_{\, ij}\, \beta_j \label{t1} \\
&& p_i= c_{\, ij}\, \beta_j + d_{\, ij}\, \alpha_j\label{t2}
\end{eqnarray}

where, ${\bf a}\ , {\bf b}\ ,{\bf c}\ ,{\bf d}$ are 
$N\times N$ transformation matrices. 
Before going into details of a particular model one needs to determine the 
conditions that the transformation matrices should satisfy. These 
conditions are obtained by the requirement of preserving the commutation
relations (\ref{comm1}),(\ref{comm2}),(\ref{comm3}) and (\ref{comm4}),
(\ref{comm5}), (\ref{comm6}). The resulting conditions are written in matrix 
form as
 
 \begin{eqnarray}
&& {\bf a}\, {\bf b}^T -{\bf b}\,{\bf a}^T = {\bf \Theta} \label{m1}\\
&& {\bf c}\, {\bf d}^T -{\bf d}\,{\bf c}^T = -{\bf B}\label{m2}\\
&& {\bf c}\, {\bf a}^T -{\bf b}\,{\bf d}^T = {\bf I}\label{m3}
\end{eqnarray}

${\bf \Theta}$ and ${\bf B}$ are anti-symmetric matrices. 
Equations (\ref{m1}), (\ref{m2}), (\ref{m3}) determine the 
 structure of the transformation  matrices. In order to illustrate in detail
 how the above procedure works, let us apply it in two dimensional space.
 As a specific model we choose a noncommutative  harmonic oscillator described 
 by the Hamiltonian
  
  \begin{equation}
H\equiv {1\over 2}\,\left[\, \left(\, p_i\,\right)^2 + \left(\, x_i\,\right)^2\,
 \right]\label{h}
\end{equation}

For simplicity we have chosen the  oscillator mass and frequency to be unity.\\
 
 Inserting (\ref{t1}), (\ref{t2}) in the 
Hamiltonian (\ref{h}) one finds its equivalent commutative form as

\begin{eqnarray}
H\equiv && {1\over 2}\, \left(\, a_{ik}\, a_{im}+d_i{}_k \, d_{im}
  \,\right)\,\alpha_k\, \alpha_m + {1\over 2}\left(\, c_{k\,i}\, c_i{}_m
  +b_{k\,i}\, b_i{}_m\,\right)\,\beta_k\,\beta_m \nonumber\\
  +&& {1\over 2} \left(\, a_i{}_k\,b_i{}_m\, +c_i{}_m \,  
  d_i{}_k \,\right)\,\left(\, \alpha_k\,\beta_m +\beta_m\, \alpha_k\,\right)
\label{hc}
\end{eqnarray}

Notice the appearance of the mixed term in conjugate coordinates and momenta
 which
is induced by  noncommutativity of the original system. 
   (\ref{m1}), (\ref{m2}), (\ref{m3}) give six conditions for determining
   matrices     ${\bf a}$,  ${\bf b}$,  ${\bf c}$, ${\bf d}$.
 On the other hand, the number of elements of the transformation matrices is 
 sixteen. $2D$ representation for ${\bf \Theta}$ and ${\bf B}$ is
  
   \begin{equation}
   \Theta_{ij}\equiv \theta\, \epsilon_{ij}\ ,\qquad
    B_{ij}\equiv B\, \epsilon_{ij}
   \end{equation}
   
   In order to solve the system one has to 
   match the number of parameters with the number of equations. One way
   to achieve this, is to assume diagonality of ${\bf a}$ and ${\bf c}$ as
   $a_{ij} \equiv a_{(i)}\,\delta_{ij}$,  
   $c_{ij}\equiv c_{(i)}\,\delta_{ij}$. With these assumptions,
    equation (\ref{m3}) imposes that diagonal elements of  matrices 
    ${\bf b}$ and  ${\bf d}$ must be zero.  Thus,  we are left with
   $8$ unknown parameters and six equations.  More equations are 
   needed. Additional equations can be obtained by requiring that the mixed 
   term in (\ref{hc}) be zero. This leads to
   
   \begin{equation}
   {\bf a}^T \, {\bf b} + {\bf d}^T\, {\bf c}=0\label{mixed}
   \end{equation}
   
   which gives  two more equations that we need. 
   The solutions of the complete set of equations can be sought by 
   the following ansatz for ${\bf b}$ and ${\bf d}$ 
   
   \begin{eqnarray}
   && b_{1\, 2}= t\, a_{2\,2}\label{b12}\\
   && b_{2\, 1}= n\, a_{1\,1}\label{b21}\\
   && d_{1\, 2}= p\, c_{2\,2}\label{d12}\\
   && d_{2\, 1}= q\, c_{1\,1}\label{d21}
   \end{eqnarray}
   
   where, we have introduced numbers $t\ , n\ ,p\ , q$  that can  take values 
   $\pm 1$ for the reason that will be clarified later on. 
   Equation (\ref{m3}) imposes condition on these numbers as
   
   \begin{eqnarray}
   && t\, p =-1\\
   && n\, q=-1
    \end{eqnarray}
    
    The complete set of solutions turns out to be 
    
   \begin{eqnarray}
   && c_{1\,1}= { 1 \over \theta}\,
   \left(\,n\, a_{1\,1} +a_{2\,2} \,\sqrt{ n\, t\, \kappa} \, \,\right)
   \label{c11}\\
   && c_{2\,2}= -{1\over \theta}\,
   \left(\, t\,  a_{2\,2} +a_{1\,1}\, \sqrt{ n\, t\,\kappa} \, \right)
   \label{c22}\\
   && a_{1\,1}^2= {\theta\over 2\,n}\,
   \left[\, 1+ {1\over\sqrt{1-4n\, t\,A^2}}\,\right]
   \label{a11}\\
   && a_{2\,2}^2= {\theta\over 2t}\,\left[\, -1+ {1\over\sqrt{1-4n\, t\,A^2}}\,
   \right]
   \label{a22}\\
   && A\equiv -{\sqrt{ n\,t\,\kappa}\over n\,t\, (\, 1+\kappa +\theta^2)}\\
   && \kappa \equiv 1 -B\,\theta
   \end{eqnarray}
    
    Inserting (\ref{c11}), (\ref{c22}), (\ref{a11}), (\ref{a22})
    in the Hamiltonian (\ref{hc}) one finds
    
    \begin{eqnarray}
    && H={1\over 2}\, \Omega_1\,\left[\, (\alpha_1)^2 + (\beta_1)^2\,\right]
    +{1\over 2}\,\Omega_2\,\left[\, (\alpha_2)^2 + (\beta_2)^2\,\right]
    \label{asym}\\
    &&\Omega_1 \equiv {1\over 2\, n}\,\left[\theta + B+\,n\, t\,
    \sqrt{ 4 + (\theta-B)^2}  \, \right]\label{fr1}\\	
    &&\Omega_2 \equiv -{1\over 2\,t}\,\left[\,\theta + B
    -n\,t\, \sqrt{ 4 + (\theta-B)^2}\, \right]\label{fr2}\\	
    \end{eqnarray}
    The Hamiltonian (\ref{asym}) is the representation of a noncommutative
    $2D$ harmonic oscillator in terms of two $1D$ commutative, anisotropic,
    harmonic oscillators. We shall name this description ``anisotropic
    representation''.
    The above result in terms of parameters $n\ ,t\ ,p\ ,q$  permits
    to consider the complete range of values of the noncommutative parameter 
    $B$, assuming $\theta >0$.
    In fact, the square root in (\ref{c11}) requires 
    $n\,t\,(\, 1-B\, \theta\,)>0$, which leads to two different ranges:
    one where $n=t=1$, $B<1/\theta$, and the other where $n=-t=1$, 
    $B> 1/\theta$. 
    Our result agrees with \cite{nair} where the two different regions 
    are described as $\kappa> 0$ and $\kappa<0$.\\
     At this point, we shall prove the existence of anther
     set of solutions for equations (\ref{m1}),
    (\ref{m2}), (\ref{m3}) which give  particularly nice representation
    of the  $2D$ noncommutative  harmonic oscillator in terms of an
    isotropic oscillator. In this representation the noncommutative effect
     have a simple and clear physical interpretation.  
    As we have already shown, the complete solution for the transformation 
    matrices can be sought starting with four independent parameters. 
    The anisotropic representation is produced by imposing diagonality
    of  ${\bf a}$ and ${\bf c}$ together with  relations 
    (\ref{b12}), (\ref{b21}), (\ref{d12}), (\ref{d21}). 
    Let us choose a different approach where the  matrices ${\bf a}$ and  
    ${\bf c}$ are kept {\it diagonal}, but with {\it single} eigenvalues $a$ 
    and $c$. The requirement of eigenvalue degeneracy reduces the
    number of free parameters by half. In order to maintain unaltered the total
    number of free parameters matrices ${\bf b}$ and ${\bf d}$,  will be 
    chosen {\it anti-symmetric}:

 \begin{eqnarray}
&& a_i{}_j\equiv a\,\delta_i{}_j \ ,\qquad c_i{}_j \equiv c\,  \delta_i{}_j\\
&& b_{ij}\equiv b\, \epsilon_{ij}\ ,\qquad d_{ij}\equiv d\, \epsilon_{ij}
\end{eqnarray}

 Equations (\ref{m1}), (\ref{m2}), (\ref{m3})  reduce to the following 
 conditions
 
 \begin{eqnarray}
 && a\, b = -{\theta\over 2} \label{ab} \\
 && c\, d = {B\over 2}\label{cd} \\
 && a\, c -b\, d = 1 \label{abcd}
 \end{eqnarray}
 
 The set of equations (\ref{ab}), (\ref{cd}), (\ref{abcd})
 enables to solve three out of four parameters as 
 
  \begin{eqnarray}
 && b=-{\theta\over 2a} \label{sol1}\\
 && c={1\over 2a}\,\left(\, 1\pm\sqrt\kappa\,\right)\ , \qquad
 \kappa\equiv 1-\theta\,B\label{sol2}\\
 && d={a\over \theta}\, \left(\, 1\mp\sqrt\kappa\,\right)\label{sol3}
 \end{eqnarray}
  
  The above solutions turn equation (\ref{m3}) into 
 
 \begin{equation}
 \left(\, \theta+B\,\right)=0\label{m33}
 \end{equation}
 
 Equation (\ref{m3})  cannot be used to determine the remaining parameter $a$, 
 as it was the case in the anisotropic representation. At most it can 
 impose relation between  parameters  $B$ and $\theta$. Our intention is to 
 work in full generality and, therefore, we shall assume $\theta + B\ne 0$.
 Thus, we shall keep the mixed term in the Hamiltonian (\ref{hc}):

 \begin{eqnarray}
 && H= h_1\, \left(\, \alpha_i\, \right)^2 + h_2\, \left(\, \beta_i\, \right)^2
 - {\theta + B\over 2}\, \epsilon_{ij}\, \alpha_i\,\beta_j 
 \label{sim}\\
 && h_1\equiv {a^2\over 2}\,\left[\, 1+{1\over \theta^2}\,\left(\, 1 
 \mp\sqrt\kappa\,\right)^2\,\right]\label{h1}\\
 && h_2\equiv {\theta^2\over 8a^2}\,\left[\, 1+{1\over \theta^2}\,
 \left(\, 1 \pm\sqrt\kappa\,\right)^2\,\right]\label{h2}
 \end{eqnarray}
 
 One can recognize (\ref{sim}) as the Hamiltonian for the  commutative, 
 isotropic, $2D$
 harmonic oscillator with {\it additional} term proportional to the
 two-dimensional angular momentum $L=\epsilon_{ij}\, \alpha_i\,\beta_j$.Thus, 
 we shall name this representation of the noncommutative $2D$ harmonic 
 oscillator ``isotropic representation''. The
 term linear in the angular momentum is the reminiscence of the 
 noncommutativity and, thus, it is important for understanding  noncommutative 
 effects. Similar term, in Quantum Mechanics, results from the coupling of the 
 angular momentum with an external magnetic field.   \\  
 Let us find out the spectrum of (\ref{sim}).  
 Schroedinger equation for the stationary states  is
 
 \begin{eqnarray}
  &&\left[ - h_2\,{\partial^2\over\partial \alpha_i^2}    
  +h_1\, \left(\, \alpha_i\, \right)^2 
  -{1\over 2}\left(\, \theta + B\,\right)\,  L\, \right]\,
 \psi\left(\,\alpha_i\,\right)= E\, \psi\left(\,\alpha_i\,\right)
 \label{realh}\\
 &&  L\equiv -i\,\epsilon_{kj}\,\alpha_k \,{\partial\over\partial\alpha_j}
  \label{ang}
 \end{eqnarray}
 
 Since Hamiltonian (\ref{realh})  is rotationally symmetric, it is
 appropriate to work in polar coordinates. In this
 coordinate system the  Schroedinger equation reads:   
 
\begin{eqnarray}
  &&\left[\, h_2\, \left(\, {1\over r}{\partial\over\partial r}r
  {\partial\over\partial r}-{  L^2\over r^2} \, 
  \right)  - h_2\,r^2 -{1\over 2}
 \left(\, \theta + B\,\right)\,  L\, \right]\,
 \psi\left(\,r\ ,\phi\,\right)= E\, \psi\left(\,r\ ,\phi\,\right)
 \label{radial}\nonumber\\
 && \\
 &&  L\equiv -i\,{\partial\over\partial\phi}
  \label{angular}
 \end{eqnarray}
 
 The Schroedinger equation  permits separation of variables in the  
 wave-function. We find it convenient to introduce the following redefinitions 
 
 \begin{eqnarray}
 &&z\equiv \sqrt{h_1\over h_2}\, r^2\\
 &&\widehat E\equiv {1\over 4\sqrt{h_1\, h_2} }\,\left[\, E- {m\over
 2}\,\left(\theta +B\right)\,\right]-{1\over 2}
 \end{eqnarray}
 
 where, we have introduced $m$ as the ``magnetic quantum number'' of the 
 angular momentum operator $L$.  The wavefunction  is of the  form
  
  \begin{equation}
 \psi(z\ ,\phi)= Z(z)\,\exp\left(-{z\over 2}+ i \, m\, \phi\,\right) 
 \end{equation}
 
 where, the radial wavefunction $Z(z)$ is subject to the equation
 
 \begin{equation}
  z\, Z^{\prime\,\prime}(z) + \left(1-z\right) Z^{\prime}(z) 
  +\left[\,\widehat E - {m^2\over 4z}\right] Z(z) =0 \label{rad}
 \end{equation}

  Equation (\ref{rad}) admits solutions in terms of Generalized Laguerre
 Polynomials as 

  \begin{eqnarray}
 && \psi_{n_r\,m}(z\ ,\phi)= N\, 
 z^{ |\,m\,|/2}\, L_{n_r}^{|\,m\,|}(z)\, \exp\left(-{z\over 2}+ i\, m\,\phi\,\right)\\
 && L_{n}^s(z)=z^{-s}\,\exp(z){d_n\over dz^n}\left(\, z^{n+s}\,\exp(-z)\,\right) 
 \end{eqnarray}

  where, $N$ is the proper normalization constant and $n_r$ is the 
  {\it radial quantum number.} The spectrum of the system is found  to be
	
	\begin{equation}
        E_{n_r\, m}= 2\sqrt{h_1\, h_2}\left(\, 2n_r +|\, m\, | + 1\,\right) 
        +{m\over 2} \left(\, \theta +B\,\right) \label{energy}
        \end{equation}

	with quantum numbers taking values $n_r= 0\ , 1\ , 2\ ,\dots$,
	$ m=0\ , \pm 1\ ,\pm 2\ ,\dots $.
   
   Using the definitions  (\ref{h1}), (\ref{h2}) one finds the frequency
   in the isotropic case to be
   
        \begin{equation}
         \omega\equiv  2\sqrt{h_1\, h_2}= {1\over 2}\, \sqrt{4 +(\theta-B)^2}
        \label{w}
        \end{equation}
  
   Equation (\ref{w}) displayes the independence of the spectrum
   on the arbitrary parameter $a$, as advocated earlier.
   The parameter $a$ simply induces a harmless (global) rescaling of the 
   radial coordinate. Let us introduce a different set of non-negative quantum 
   numbers $n_+ \ , n_-$ \cite{messhia} in terms of which the radial and 
   magnetic quantum numbers are written as 
   
   \begin{equation}
   n_r\equiv n_- +{m-|\,m\,|\over 2} \ ,\qquad m\equiv n_+ - n_-
   \end{equation}
   
   and the spectrum turns out to be
   
   \begin{equation}
    E_{n_+\, n_-}= \omega\left(\, n_+ + n_- + 1\,\right) +{ n_+ - n_-\over 2}
    \left(\, \theta +B\,\right) \label{e+-}
    \end{equation}
    
   The spectrum (\ref{e+-}) in the special case $B=0$ has been studied 
   in \cite{gamb}.\\
   For the sake of transparency, we  write down 
   the first two excited states of the noncommutative harmonic oscillator 
   
   \begin{eqnarray}
   &&E_{\, 0 ,2}= 3\omega + \left(\, \theta +B\,\right)\\ 
   &&E_{\, 1 ,1}= 3\omega \\
   &&E_{\, 2 ,0}= 3\omega -\left(\, \theta +B\,\right)\\
   &&\nonumber\\
   &&E_{\, 0 ,1}= 2\omega +{1\over 2} \left(\, \theta +B\,\right)\\ 
   &&E_{\, 1 ,0}= 2\omega -{1\over 2} \left(\, \theta +B\,\right)\\
   &&\nonumber\\
   &&E_{\, 0 ,0}= \omega
   \end{eqnarray}
   
   We point out that the spectrum  displays the fact that  the noncommutativity
   parameters play the same role as an external magnetic field ${\cal H}\equiv
   \theta + B$.  This role of the noncommutative parameters explains the choice 
   made in \cite{greci} as corresponding to the absence 
   of the ``{\it magnetic field}''  and, thus, re-establishing energy levels
   degeneracy.
   
   \section{Discussion and conclusions}
  
   As we have shown, the sets of solutions (\ref{sol1}),
   (\ref{sol2}), (\ref{sol3}) and (\ref{c11}) ,  (\ref{c22}), (\ref{a11}),
   (\ref{a22}) lead to two representations of the noncommutative harmonic 
   oscillator.
   We would like to show that the spectra of the two modes are identical. 
   Let us re-write (\ref{e+-}) as
   
   \begin{equation}
    E_{n_+\, n_-}=  
    \left(\, \omega +\theta +B\,\right)\left(\,  n_+ + {1\over 2}\,\right)
    +\left(\, \omega -\theta -B\,\right)\left(\,  n_- + {1\over 2}\,\right)
    \label{e+-2}
    \end{equation}
   
   The energy spectrum (\ref{e+-2}) matches the one of the
   Hamiltonian  (\ref{sim}), provided one identifies parameters as follows
   
   \begin{eqnarray}
  && \omega= {1\over 2}\left(\, \Omega_1 + \Omega_2\, \right)\\
   && \theta +B= {1\over 2}\left(\, \Omega_1 - \Omega_2\, \right)
   \end{eqnarray}
   The advantage of 
    the representation in terms of isotropic oscillator is that it offers  
    clear identification for the noncommutativity as magnetic field effect. 
    On the other hand, the equivalence to the anisotropic representation 
    displayes that the effect of the magnetic field can be simulated by the 
    frequency difference of the anisotropic oscillators. Similar conclusion,
    in a different context and in terms of chiral oscillators has been found 
    in \cite{baner}.\\ 
    The equivalence of the spectra displays that the two descriptions of the
    noncommutative harmonic  oscillator are equivalent, in spite of the
    asymmetry with respect to rotations.  The generator of  rotations, i.e. 
    angular momentum, is defined by the commutators 
  
    \begin{eqnarray}
    &&\left[\, L\ , \alpha_k\,\right]= i\,\epsilon_{kj}\,\alpha_j\, \label{c1}\\
    && \left[\, L\ , \beta_k\,\right]= i\,\epsilon_{kj}\,\beta_j \label{c2}
    \end{eqnarray}
    
    Therefore, our definition of isotropic representation is motivated by
    the fact that the Hamiltonian (\ref{realh}) commutes with $L$.
    Let us express the angular momentum operator (\ref{ang}) in terms 
    of the noncommutative coordinates $(\, x\ , p\,)$  with the help of
    (\ref{sol1}),   (\ref{sol2}),  (\ref{sol3}).  One finds that the 
    noncommutative form of $L$, call it $J$, is

    \begin{equation}
    J={1\over \kappa }\,\left(\, \epsilon_{ij}\, x_i\, p_j + {\theta\over 2}\,
    p_i^2
    + {B\over 2}\, x_i^2\,\right)\label{j}
    \end{equation}
    
    The additional terms take into account noncommutativity in $\theta$ and $B$.
   (\ref{j}) has the form found in (\cite{nair}). $J$  remains the angular 
   momentum in the space of noncommutative coordinates. In fact, it satisfies 
    
    \begin{eqnarray}
 && \left[\, J\ ,x_k\,\right]= i\,\epsilon_{kl}\,x_l\\
 && \left[\, J\ ,p_k\,\right]= i\,\epsilon_{kl}\,p_l\\
     \end{eqnarray}

    In both forms the angular momentum satisfies
    $\left[\, H\ , J\, \right]=\left[\, H\ , L\, \right]=0$. 
    Thus, commutative ( in terms of $\alpha$ and $\beta$ ) and
    noncommutative ( in terms of $x$ and $p$ )  representations are isotropic. 
    \\
    Let us return, now, to the anisotropic representation.
    We shall name the canonical coordinates  of this representation
    $Q_i$, $P_i$ for easier distinction from other representations.
    Rotations are still generated by the angular momentum operator 
    $L= \epsilon_{ij}\, Q_i\, P_j$ satisfying relations (\ref{c1}), (\ref{c2}). 
    On the other hand, one can verify that $\left[\, H\ , L\,\right]\ne 0$ 
    implying the absence of rotational symmetry. 
    The  isotropic and anisotropic representations have different symmetry
    properties with respect to rotations, and  the rotational symmetry cannot 
    be responsible for the equivalence of their spectra.
    We would like to identify the symmetry which leaves the spectra unchanged.
    For this purpose, let us express the angular momentum operator of the 
    isotropic representation (\ref{ang}) in terms of the coordinates $Q_i$ and 
    $P_i$. We write down the relation among coordinates of the two 
    representations 
    
    \begin{eqnarray}
    &&\alpha_1= {1\over \sqrt 2}\, \left({h_2\over h_1}\right)^{1/4}\left(\,
    Q_1-P_2\,\right)\\
   &&\alpha_2= -{1\over \sqrt 2}\, \left({h_2\over h_1}\right)^{1/4}\left(\,
    Q_2-P_1\, \right)\\
   &&\beta_1= {1\over \sqrt 2}\, \left({h_1\over h_2}\right)^{1/4}\left(\,
   Q_2+P_1\, \right)\\
   &&\beta_2 = -{1\over \sqrt 2}\, \left({h_1\over h_2}\right)^{1/4}\left(\,
   Q_1+ P_2\, \right)
   \end{eqnarray}
    
    which give the form of $L$ in (\ref{ang}), call it now  $\widehat L$, as 
    
    \begin{equation}
     \widehat L={1\over 2}\, \left(\, Q_2^2+ P_2^2-Q_1^2-P_1^2\, \right)
     \label{lhat}
     \end{equation}

     One finds 
    
    \begin{eqnarray}
    &&\left[\,\widehat L\ , Q_i^2\,\right]= 2i\,\left[\, Q_1 P_1 - Q_2 P_2\,
    \right]
    \label{chat1} \\
    && \left[\,\widehat L\ , P_i^2\,\right]= 
    -2i\,\left[\, Q_1 P_1 - Q_2 P_2\,\right]
    \label{chat2}\\
    \end{eqnarray}
    
    which show that (\ref{lhat}) is not the generator of   rotations.
    On the other hand, one finds that $\left[\,\widehat L\ , H\, \right]=0$. 
    It is possible to verify that 
    
    \begin{equation}
    \left[\,\widehat L\ , L\,\right]= -2i \,\overline L\equiv 2i\, 
    \left(\, Q_1 Q_2 + P_1 P_2\,\right)
    \end{equation}
    
     and 
    
    \begin{eqnarray}
    &&\left[\,\overline L\ , Q_1^2\,\right]= 2i\, Q_1\, P_2 
    \label{chat3} \\
    && \left[\,\overline L\ , Q_2^2\,\right]= 2i\, Q_2\, P_1 
    \label{chat4} \\
    && \left[\,\overline L\ , P_1^2\,\right]= -2i\, Q_2\, P_1 
    \label{chat5}\\
    && \left[\,\overline L\ , P_2^2\,\right]= -2i\, Q_1\, P_2 
    \label{chat6}\\
    \end{eqnarray}
    
    One realizes that the operators $L$, $\widehat L$,
    $ \overline L$ form an $SU(2)$ algebra \cite{bellucci}.
    The equivalence of the spectra in the anisotropic
    and isotropic representations must, therefore, be result of the invariance 
    of the Hamiltonian with respect to the above $SU(2)$ group.
    To put in evidence the $SU(2)$ invariance of the Hamiltonian  let us 
    calculate the sum of the squares of the three operators $L$, $\widehat L$, 
    $\overline L$ in the anisotropic representation. One finds

     \begin{equation}
     L^2 + \widehat L^2 + \overline L^2 =
     {1\over 4}\left[\, Q_i^2 + P_i^2\,\right]^2 -1 \equiv C^2 -1
     \end{equation}
    
    The result shows the existence of the operator $C$ which permits
    the following expression of the Hamiltonian (\ref{hc}) 
     
     \begin{equation}
    H={ \Omega_1 + \Omega_2\over 2}\, C -  {\Omega_1 -\Omega_2\,\over 2}\, 
    \widehat L
    \label{hgen}
    \end{equation}
    
    Expression (\ref{hgen}) indicates that the  set of commuting  
    operators, needed to describe the spectrum, is $H$, $C$, $\widehat L$ which
    also exhibits invariance of the Hamiltonian under $SU(2)$.
   The eigenvalues of these operators can be described in terms of two quantum 
   numbers  $n_2$ and $n_1$, associated to the operators  $Q_2^2+ P_2^2$,
   $Q_1^2+ P_1^2$ respectively. One can verify that this description leads to
   the spectrum (\ref{e+-2}) with $n_1 =n_+$,  $n_2 =n_-$. In passing from the 
   anisotropic to the isotropic  representation, one rewrites
   operators $C$ and $\widehat L$ in terms of the  coordinates of the 
   appropriate representation. In doing so, $\widehat L$ becomes the angular
    momentum operator in the isotropic representation re-establishing 
    rotational symmetry. One can verify that (\ref{hgen}) reproduces (\ref{sim})
    and (\ref{h}).\\
    
    In this paper we have shown the existence of an isotropic representation
    of the noncommutative harmonic oscillator which goes hand-by-hand with,
    already known, anisotropic representation. These two representations are
    different if seen from the point of view of rotational symmetry. The
    reason for this symmetry breaking can be traced back to the choice of the
    ansatz for the transformation matrices connecting commutative and
    noncommutative coordinates. Different eigenvalues of the  transformation
    matrices break rotational symmetry explicitly. Nonetheless, the two
    representations describe the same physical system,  as the
    equivalence of the spectra show. 
    The symmetry of the spectrum, for any choice of the ansatz, is the $SU(2)$ 
    symmetry described in the discussion above. 
    The isotropic representation has the advantage of giving clear physical
    meaning to the effect of noncommutativity as being equivalent to an external
    magnetic field.  There may be other representations  of
    the noncommutative system corresponding to different  solutions for
    the transformation matrices, but they should all lead to one of the
    two forms of the Hamiltonian described in this paper.\\
    Finally, we would like to correct generally
    accepted, but not completely appropriate, use of the term magnetic field 
    when referring ,{\it only}, to  the noncommutative parameter $B$. This is
    motivated by the fact that the noncommutative momentum turns into
    a covariant one in terms of canonical coordinates. However, the role
    of coordinates and momenta is equivalent in phase space, and thus
    it is  clear that the parameter $\theta$ plays  the same role as $B$. 
    This is displayed in (\ref{sim}). Therefore, parameter $\theta$
    equally deserves the name of ``magnetic field'' \cite{baner}. 
   
   \begin{acknowledgments}
   We would like to thanks Prof. A.Jellal and Prof. R.Banerjee for useful
    discussions, and Prof. M.Crescimanno for providing us a copy of the original
   paper by V.Fock where the $2D$ harmonic oscillator coupled to a magnetic
   field was quantized \cite{fock}.
   \end{acknowledgments}


\begin{thebibliography}{99}
\bibitem{witten}  E. Witten
Nucl. Phys. B\textbf{460} 335  (1996) 
\bibitem{sw} N.Seiberg,  E. Witten
JHEP \textbf{9909}  032 (1999)
\bibitem{schw} A. Konechny, A. Schwarz
 {\it  Introduction to M(atrix) theory and noncommutative geometry, Part II}
  hep-th/0107251\\
 A. Konechny, A. Schwarz
 {\it  Introduction to M(atrix) theory and noncommutative geometry}
  hep-th/0012145
\bibitem{nekra}M. R. Douglas, N. A. Nekrasov
 {\it Noncommutative Field Theory};  hep-th/0106048
\bibitem{presn} M. Chaichian, A. Demichev, P. Pre\v snajder
Nucl. Phys. B\textbf{567 } 360   (2000)
\bibitem{agw}L.Alvarez-Gaume, S. R. Wadia
 Phys. Lett. B \textbf{501} 319 (2001)\\
 L. Alvarez-Gaume, J.L.F. Barbon 
 Int. J. Mod. Phys.\textbf{A16} 1123  (2001) 
\bibitem{qm} 
  R. P. Malik, A. K. Mishra, G. Rajasekaran
   Int. J. Mod. Phys. \textbf{A13}4759 (1998)\\
  V.P. Nair 
  Phys. Lett. B \textbf{505} 249 (2001) \\
   B. Morariu, A. P. Polychronakos
  {\it Quantum Mechanics on the Noncommutative Torus}; hep-th/0102157
 \bibitem{nair} V.P. Nair, A.P. Polychronakos 
Phys. Lett. B \textbf{505}  267 (2001)
 \bibitem{greci} A. Hatzinikitas, I. Smyrnakis {\it  The noncommutative
 harmonic
oscillator in more than one dimensions}, hep-th/0103074
 \bibitem{gamb}J. Gamboa, M. Loewe, F. Mendez, J. C. Rojas
 {\it The landau problem and noncommutative Quantum Mechanics};
 hep-th/0104224
 \bibitem{jell}A. Jellal
{\it Orbital Magnetism of Two-Dimension Noncommutative Confined System}
hep-th/0105303\\
O.F. Dayi, A. Jellal
 {\it Landau Diamagnetism in Noncommutative Space and the Nonextensive 
 Thermodynamics of Tsallis;} cond-mat/0103562 
 \bibitem{bellucci}  S.Bellucci, A. Nersessian, C.Sochichiu 
{\it Two phases of the noncommutative quantum mechanics}, hep-th/0106138
\bibitem{...}J. Gamboa, M. Loewe, F. Mendez, J. C. Rojas
 {\it Noncommutative Quantum Mechanics: The Two-Dimensional Central Field};
 hep-th/0106125 
 \bibitem{uno}
 The dimensions  of various noncommutative constants 
are $\left[\, \Theta\, B\,\right]=\left[\,\hbar^2\,\right]$. For simplicity
we choose $\hbar=1$.
\bibitem{messhia} A. Messiah {\it Quantum Mechanics}, Ch.XII, North Hollad
Publ. Comp. (1961)
\bibitem{baner} R. Banerjee
Phys. Rev.D\textbf{60}   085005  (1999)\\
R. Banerjee
{\it Dissipation and Noncommutativity in Planar Quantum Mechanics}
hep-th/0106280
 \bibitem{fock}V. Fock
"Bemerkung zur Quantelung des HO im Magnetfeld"
Zeitschrift F\"ur Physik  \textbf{84} 446 (1928)
\end{thebibliography}
\end{document}